\documentclass[preprint,12pt]{aastex}

\input{epsf}

\def\plotfiddle#1#2#3#4#5#6#7{\centering \leavevmode
\vbox to#2{\rule{0pt}{#2}}
\includegraphics{#1}}

\newcommand{\ks}{K$_s$}

\slugcomment{Version 0.99; 31 Jan 2007}

\begin{document}


\title{The Spitzer {\it c2d} Survey of Large, Nearby, Interstellar Clouds\\
    VIII.  Serpens Observed with MIPS}


\author{Paul M. Harvey\altaffilmark{1}, Luisa M. Rebull\altaffilmark{2}, 
Tim Brooke\altaffilmark{3}, 
William J. Spiesman\altaffilmark{1}, 
Nicholas Chapman\altaffilmark{4}, 
Tracy L. Huard\altaffilmark{5}, Neal J. Evans II\altaffilmark{1}, 
Lucas Cieza\altaffilmark{1},
Shih-Ping Lai\altaffilmark{4},
Lori E. Allen\altaffilmark{5},
Lee G. Mundy\altaffilmark{4},
Deborah L. Padgett\altaffilmark{2},
Anneila I. Sargent\altaffilmark{3},
Karl R. Stapelfeldt\altaffilmark{6}
Philip C. Myers\altaffilmark{5},
Ewine F. van Dishoeck\altaffilmark{7},
Geoffrey A. Blake\altaffilmark{8},
David W. Koerner\altaffilmark{9}
}

\altaffiltext{1}{Astronomy Department, University of Texas at Austin, 1 University Station C1400, Austin, TX 78712-0259;  pmh@astro.as.utexas.edu, nje@astro.as.utexas.edu, lcieza@astro.as.utexas.edu, spies@astro.as.utexas.edu}
\altaffiltext{2}{Spitzer Science Center, MC 220-6, Pasadena, CA 91125; rebull@ipac.caltech.edu; dlp@ipac.caltech.edu}
\altaffiltext{3}{Division of Physics, Mathematics, \& Astronomy, MS 105-24, California Institute of Technology, Pasadena, CA 91125; tyb@astro.caltech.edu; afs@astro.caltech.edu}
\altaffiltext{4}{Astronomy Department, University of Maryland, College Park, MD 20742; chapman@astro.umd.edu, slai@astro.umd.edu, lgm@astro.umd.edu}
\altaffiltext{5}{Smithsonian Astrophysical Observatory, 60 Garden Street, MS42, Cambridge, MA 02138; leallen@cfa.harvard.edu
.edu, thuard@cfa.harvard.edu, pmyers@cfa.harvard.edu}
\altaffiltext{6}{Jet Propulsion Laboratory, MS 183-900, California Institute of Technology, 4800 Oak Grove Drive, Pasadena, CA 91109; krs@exoplanet.jpl.nasa.gov}
\altaffiltext{7}{Leiden Observatory, Postbus 9513, 2300 RA Leiden, Netherlands; ewine@strw.leidenuniv.nl}
\altaffiltext{8}{Division of Geological and Planetary Sciences, MS 150-21, California Institute of Technology, Pasadena, CA 91125; gab@gps.caltech.edu}
\altaffiltext{9}{Northern Arizona University, Department of Physics and Astronomy, Box 6010, Flagstaff, AZ 86011-6010; koerner@physics.nau.edu}


\begin{abstract}

We present maps of 1.5 square degrees of the Serpens dark cloud at 24, 70, and 160\micron\ observed
with the Spitzer Space Telescope MIPS Camera.  We describe the observations and briefly discuss the data processing
carried out by the c2d team on these data.  More than 2400 compact sources have been extracted at 24\micron, nearly 100 at 70\micron,
and 4 at 160\micron.
We estimate completeness limits for our 24\micron\
survey from Monte Carlo tests
with artificial sources inserted into the Spitzer maps.
We compare source counts, colors, and magnitudes in the Serpens cloud to
two reference data sets, a 0.50 deg$^2$ set on a low-extinction region near the dark cloud, and a
5.3 deg$^2$ subset of the SWIRE ELAIS N1 data that was processed through our pipeline.  
These results show that there is an easily identifiable population of young stellar
object candidates in the Serpens Cloud that is not present in either of the reference data sets.
We also show a comparison of visual extinction and cool dust emission illustrating a close
correlation between the two, and find that the most embedded YSO candidates are located in the
areas of highest visual extinction.

\end{abstract}


\keywords{infrared: general --- clouds: star forming regions}



\section{Introduction}
The Spitzer Space Telescope Legacy project ``From Molecular Cores to Planet-forming Disks'' includes
IRAC and MIPS mapping of five large star-forming clouds \citep{evans03}.  The Serpens cloud covers more
than 10 square degrees as mapped by optical extinction \citep{camb99}, but for
reasons of practicality the c2d project was only able to observe 1.5 deg$^2$ with the
MIPS instrument on Spitzer (further Spitzer observations of a larger area of Serpens
are planned as part of an extended survey of the Gould Belt, Allen 2007, in prep.).
At an assumed distance of 260 pc \citep{stra96}, the area mapped by c2d corresponds to $\sim$ 4.5 $\times$ 7 pc.
This paper is one in a series describing the IRAC and MIPS observations of each of the
c2d clouds.  Previous papers include those on IRAC observations of Serpens \citep{har06},
Chamaeleon \citep{porras07},
and Perseus \citep{jorg06}, as well as MIPS observations of Chamaeleon \citep{young05},
Perseus \citep{rebull07}, Lupus \citep{chapman07}, and Ophiuchus \citep{padgett07}.

Our observations of Serpens cover an area that includes the well studied ``core''
cluster region, Cluster A, together with the newly discovered Cluster B \citep{har06,djup06} to the south,
as well as the Herbig Ae/Be star, VV Ser.  Significant portions of this cloud have been
studied by previous space infrared missions, including IRAS \citep{zhang88a,zhang88b} and
ISO \citep{kaas04,djup06}.  The much higher sensitivity and longer wavelength capability of
the Spitzer MIPS instrument, however, allows us to detect both very low luminosity infrared-excess
objects and to map very cool diffuse dust emission in the region.  Our results are also complementary
to the 1.1mm mapping of the same region by \citet{enoch07}.  The combined results on Serpens using
both the MIPS and IRAC observations are discussed in a companion paper where we also give detailed
object lists \citep{har07}.

In \S \ref{obs} we describe details of the observations obtained from the MIPS instrument for
Serpens and the data processing pipeline used to reduce the observations. 
In \S\ref{results} we
describe a number of results from our MIPS observations and correlations between them and
the 2MASS catalog \citep{skrut06}.  
We show in \S\ref{diffuse}
that there is an excellent correlation between the coolest dust that we can observe which emits
at 160\micron\ and the optical extinction in Serpens.
We investigate the
possibility of time variability at 24\micron\ in our two-epoch data set in \S\ref{var}.
In \S\ref{counts}
we discuss our results statistically in terms of source counts and compare these to predictions
of models of the Galaxy as well as to the counts in the reference fields.  
We present color-color and color-magnitude plots of the population of
infrared sources in \S\ref{color} and discuss the separation of likely cloud members from the
extensive background population of stars and extragalactic objects.  In the final part of \S\ref{results}
we briefly describe some details of
individual sources of particular interest.

\section{Observations and Data Reduction}\label{obs}

\begin{figure}
\epsscale{1.0}
\plotone{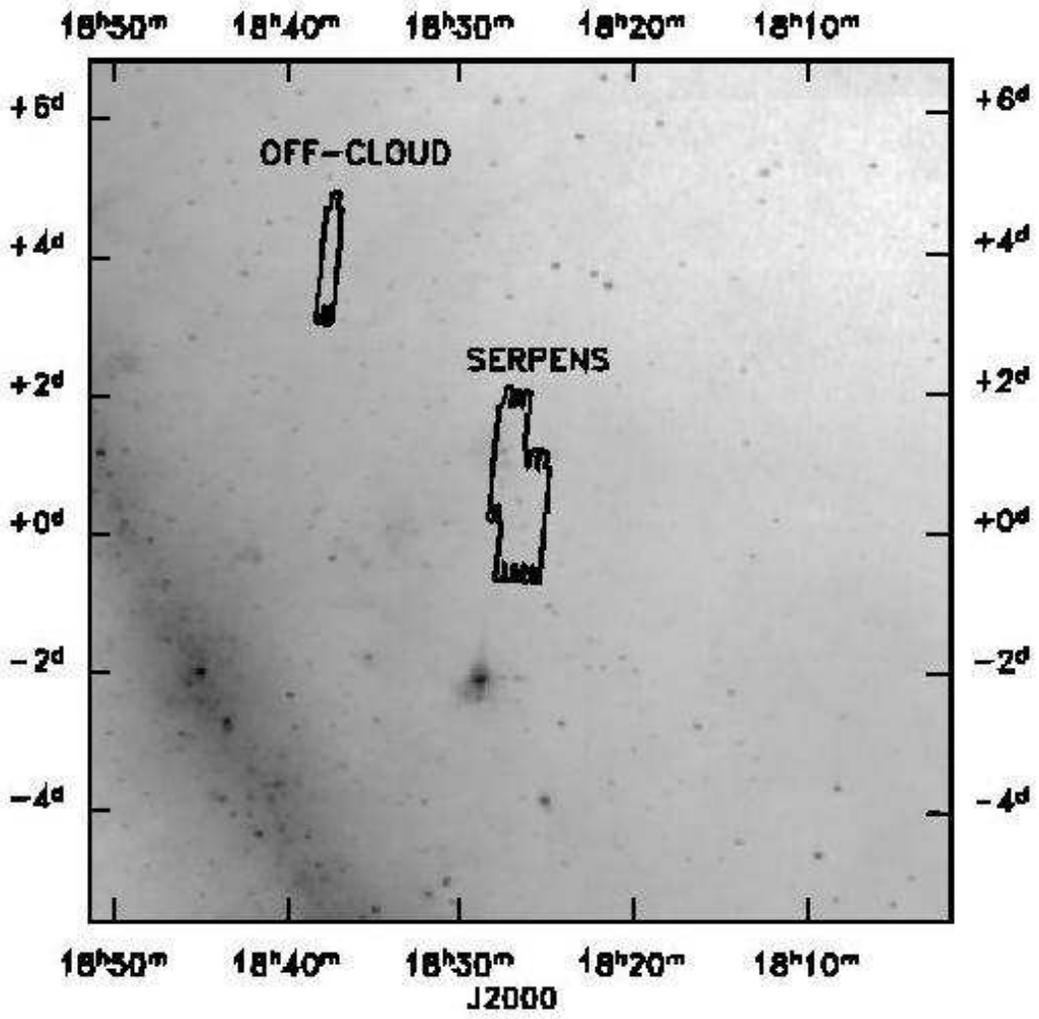}
\caption{IRAS 25 $\mu$m map showing the observed {\it c2d} regions in
the Serpens cloud, both the star-forming region marked ``SERPENS'' and the low-extinction
``OFF-CLOUD'' area.
\label{fig_iras25}}
\end{figure}

The MIPS observations cover an area where $A_v > 6$ in the contour maps of
\citet{camb99}.  In addition, a nearby off-cloud region of 0.5 square
degrees was mapped for comparison with the cloud region.  
A summary of the
regions observed is listed in Table \ref {table_summary} with the AOR (Astronomical Observation Request)
number to facilitate access from the Spitzer archive.
The regions covered at 24\micron\ are outlined in Figure \ref {fig_iras25} against the 25 $\mu$m
IRAS sky.
The observing strategy and basic MIPS data analysis for the c2d star-forming clouds have been
described in detail by \citet{rebull07}, but we summarize here the most important
details.  Fast scan maps were obtained at two separate epochs with a spacing between adjacent
scan legs of 240" in each epoch.  The second epoch observations were offset by 125" from the first in
the cross-scan direction to fill in the 70\micron\ sky coverage that would otherwise have been missed
due to detector problems.  The second epoch scan was also offset 80" from the first in the scan
direction to minimize missing 160\micron\ data.  For some of the c2d clouds, these offsets together with sky rotation were
sufficient to give essentially complete one-epoch coverage at 160\micron, but for Serpens there
were still small gaps between every two scan lines.
Table \ref {table_coverage} lists the sky coverage at each wavelength.
The two observation epochs were separated in time by $\sim$ 6 hours to allow identification of
asteroids in the images; over this time period asteroids will typically move 0.3 -- 2 arcminutes.
Because of Serpens' relatively large ecliptic latitude, $\sim$ 24 degrees, only a very small number of asteroids were
seen, all of which were removed by requiring 2-epoch detection in our final source lists.
Typical 
integration times are 30 seconds at 24\micron, 15
seconds at 70\micron\ and 3 seconds at 160\micron.
Additional GTO observations east of the region of highest emission are not
included in this analysis because a different observing strategy was used.
Those observations could, however, be added to ours in order
to construct a somewhat larger mosaic of the region.

\begin{figure}
\epsscale{1.0}
\plotone{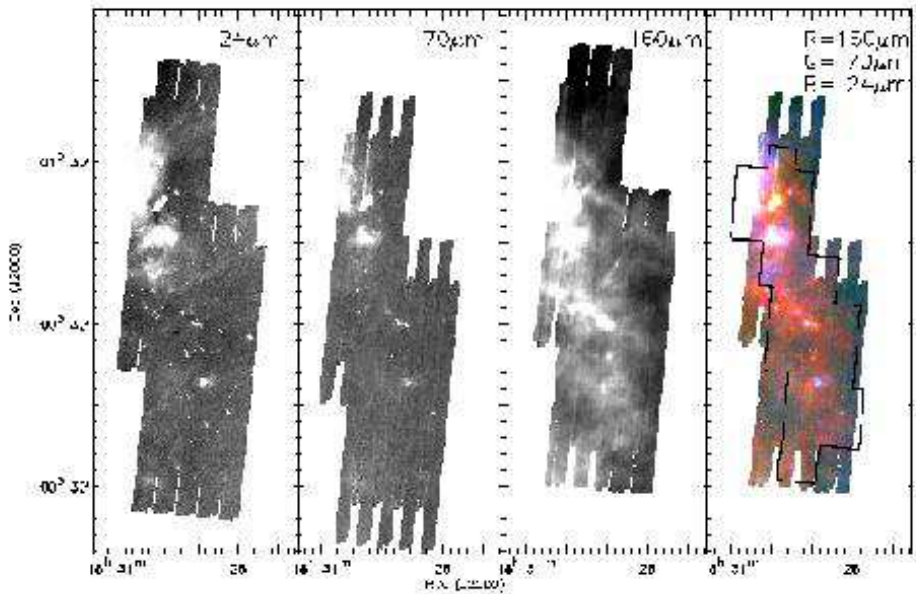}
\caption{Registered Serpens 24\micron, 70\micron\ and 160\micron\
images of the {\it c2d} MIPS region. The color image is a composite
of all three bands, and includes only the 1.27 square degree area where
data are available for each of the three bands.  Colors represent
red:160\micron\ green:70 \micron\ and blue:24\micron.
The black outline shows the region where 4 bands of IRAC data were observed \citep{har06}.
\label{fig_panel4}}
\end{figure}

Figure \ref{fig_panel4} shows the three individual images produced for the MIPS bands as well as a false
color image of the three together.  \citet{har07} show an additional image combining the
24\micron\ data with IRAC observations as well as enlargements of the two main clusters observed.
Note that unlike the IRAC instrument, the three wavelengths of MIPS all have diffraction limited
spatial resolution which means the resolution varies dramatically between 24\micron\ ($\sim$ 6") and
160\micron\ ($\sim$ 40").

Our data reduction is described in detail by \citet{evans07} but we summarize the important
details here.  In addition, previous versions of the c2d pipeline, some of which still apply to
these data, have been described in more detail by \citet{rebull07} and  \citet{young05}.  We began our data reduction with the BCD images, processed
in this case by the standard SSC S13.2 pipeline.  Following this the three MIPS channels underwent
slightly different processing paths in our c2d reduction.  The 24\micron\ data 
were mosaicked with the SSC's Mopex software \citep{mak05} after processing in the c2d pipeline
to reduce  artifacts, e.g. ``jailbars'' near bright sources.  Point sources were extracted
with ``c2dphot'' (Harvey et al. in prep.), a source extractor based on ``Dophot'' \citep{dophot},
which utilizes the mosaics for source identification but the stack of individual BCD's for
each identified object to provide the photometry and position information.
We have estimated our completeness limit at 24\micron\ in a manner similar to that described
for our IRAC photometry \citep{har06}.  We inserted a number of artificial sources into the
24\micron\ mosaic at random positions over a range of brightness covering the range $2 <$ [24] $< 12$ mag.
and then tested whether they were properly extracted.   We also produced a mosaic with
only artificial sources (no real ones) but a noise level comparable to that in the observed image, and tested the
completeness of extraction from that artificial image to estimate the effects of confusion
in this relatively high source density region.  Figure \ref{complete} shows the results from
these tests.  Clearly at the fainter flux levels, the effects of high source density are
important to the true completeness level in Serpens, e.g. [24] $> 9.5$ mag.

The processing of the 70\micron\ data followed a path similar to that at 24\micron\ with
two exceptions.  At 70\micron\ the SSC produces two sets of BCD's, one of which is simply
calibrated and another that is filtered spatially and temporally in a manner that
makes point source identification easier but which does not conserve flux for brighter
sources nor for diffuse emission.  We produced mosaics of both the unfiltered and
the filtered products using Mopex on the native BCD pixel scale.
Point
sources were extracted  using APEX \citep{mak05}.  
Source reality
was checked by hand inspection and comparison with the 24\micron\ source list.
Generally the filtered mosaics were used for point source extraction, but above F(70) $\sim$ 2 Jy,
we used the unfiltered data.
Above F(70) $\sim$ 23 Jy, sources begin to be saturated.  At these very high flux levels we used a procedure
to fit the wings of the source profile; these data have been assigned a higher uncertainty of 
because of the inherent uncertainties in this procedure.

Complete tables of source positions and flux densities for likely cloud members in Serpens are given
by \citet{har07} for our 3.6 -- 70\micron\ observations.
At 160\micron\ our processing was limited to producing a native pixel scale mosaic using
interpolation to fill in missing pixels and point source extraction from the unfiltered mosaic.
We extracted four nominal point sources in the entire mapped area.  Two of these
are associated with obvious multiple clumps of 24/70\micron\ sources.
The other two,
SSTc2dJ1829167+0018225 (associated with IRAS 18267+0016) and SSTc2dJ18293197+0118429 (associated with
source 159 of \citet{kaas04}) are likely powered mostly by single, shorter wavelength sources.
Table \ref{tbl_160} lists the positions and flux densities of these four nominal point sources
with short comments, since their 160\micron\ photometry is not described in any of our other publications on Serpens.
None of these is in the core area of either of the main clusters.  This is because large areas in
those clusters are saturated, and
the close spacing of many bright sources
leads to the complicated, extended structure seen in Figure \ref{fig_panel4} at 160\micron, without
obvious point-like sources.

After extraction, the source lists were bandmerged with our IRAC source lists
for Serpens \citep{har06} and the 2MASS catalog of J, H, and K$_s$ photometry
\citep{skrut06} as described by \citet{evans07}.  The radius for source matching
with shorter wavelength detections was 4" at 24\micron\ and 8" at 70\micron.
Table \ref{src-counts} lists the number of sources extracted at 24 and 70\micron,
and some examples of statistics of numbers identified with shorter wavelength
sources.  In addition to bandmerging, sources undergo a classification process
based on the available photometry, 2MASS, IRAC, and MIPS.  For the purposes of
this paper the most important classification is that of ``star'' which implies
a spectral energy distribution that is well-fit as a reddened stellar
photosphere without requiring any excess infrared emission from possible
circumstellar dust.  The data reported here consist of a subset of all the sources
extracted in Serpens.  The entire catalog is available from the SSC website
(http://ssc.spitzer.caltech.edu/legacy/all.html).  For this paper we have limited our discussion to sources
with a signal-to-noise ratio greater than 5 and to sources found in both
epochs of observation to eliminate asteroids.  These limits lead to a very high
reliability for the objects reported here, probably greater than 98\%.

In addition to our reduction of the Serpens Cloud and off-cloud data, we have
also processed a 5.3 deg$^2$ portion of the SWIRE Spitzer Legacy data \citep{swire} from the
ELAIS N1 field through our c2d pipeline.  Since this field is almost entirely
populated by Galactic stars and extragalactic objects, it provides an additional
control field against which to compare our Serpens Cloud population as discussed
below.  Note that the SWIRE observations go approximately a factor of 4 deeper
than c2d due to increased integration time.

\section{Results}\label{results}

\subsection{Extended Emission}\label{diffuse}

\begin{figure}
\epsscale{0.6}
\plotone{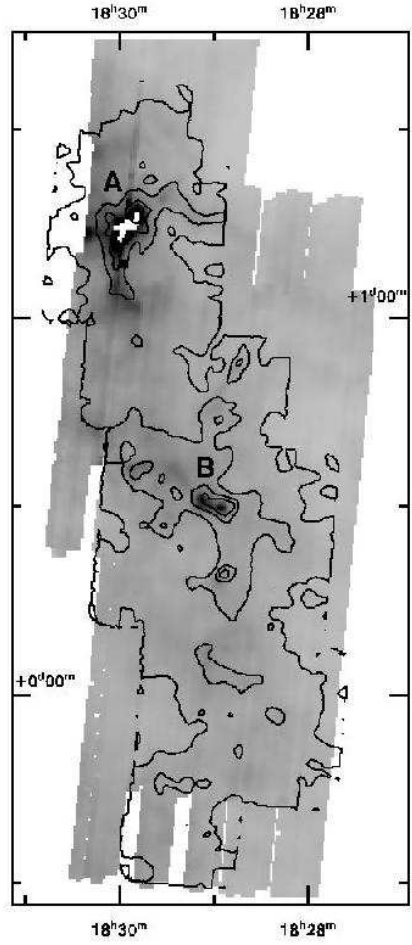}
\caption{Contours of $A_v$ at levels of 5,10,20,30 mag determined from 2MASS and Spitzer c2d IRAC
data are overlaid on the Serpens 160 $\mu$m image.  The visual extinction  and
160$\mu$m emission are quite well correlated. The locations of Cluster A and B are indicated.
\label{fig_avimage}}
\end{figure}

The 160\micron\ emission traces the coolest and most extended dust seen with MIPS.
Figure \ref{fig_avimage} shows an image of the 160\micron\ emission together with
contours of the optical extinction.  Also shown are the locations of the two main
clusters of young stellar objects in Serpens, the core Cluster A, and Cluster B
(also called the G3-G6 cluster by \citet{djup06}).  The optical extinction has
been estimated by our fitting of the objects that were well characterized as
extincted stellar photospheres. This figure shows a very close
correlation between the coolest dust and the dust that is associated with optical
extinction.  The figure also clearly shows that the two high-stellar-density clusters,
Cluster A and B,
are located in areas of maximum extinction, as we discuss further in \S \ref{embed}.

\subsection{24\micron\ Time Variability}\label{var}

\begin{figure}
\epsscale{1.0}
\plotone{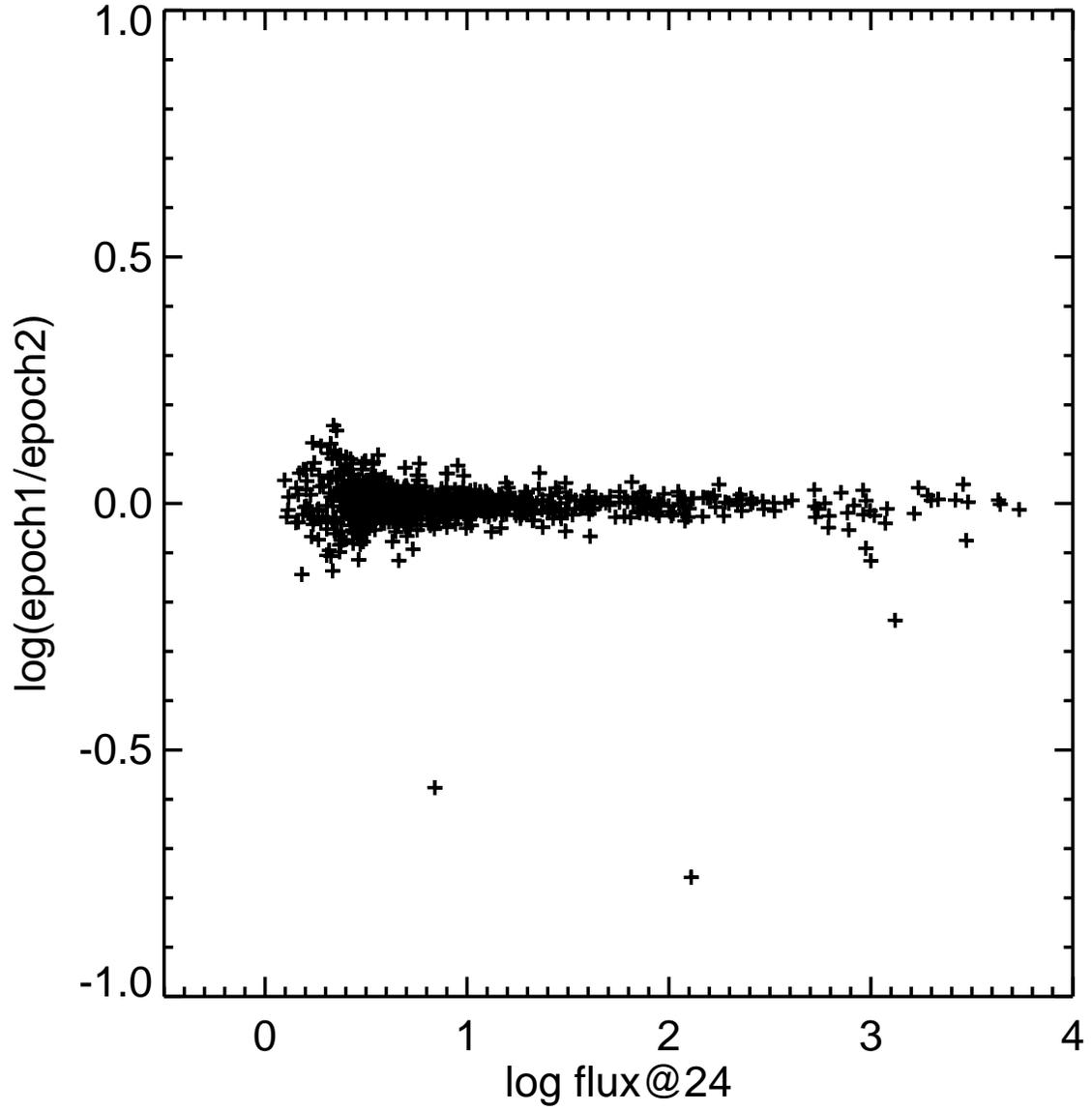}
\caption{A search for time variability in the Serpens 24\micron\ data; plot of log flux ratio of epoch1 to epoch2
versus log flux density (mJy) for the combined epoch data.  There is
no verifiable time variable source in the cloud based on these data.
\label{time_var}}
\end{figure}

Since many pre-main-sequence stars exhibit variable optical emission, we conducted a simple
examination of the 24\micron\ fluxes from the two observed epochs, similar to
that in Perseus by \citet{rebull07} and for the IRAC data in Serpens \citep{har07}.
As shown in Table \ref{table_summary}, the time difference between the two epochs of
observation was of order 4 hours.
Figure \ref{time_var} shows the ratio of the 24\micron\ flux density between the two epochs for all the
extracted sources whose signal-to-noise ratio was above 5 that were detected in
both epochs of observation.  Although there are a few outliers beyond the limits
expected on the basis of the signal-to-noise ratios, these are all readily explained
as due to poor photometry near the edges of the mosaic or problems due to source confusion
or adjacency to
bright sources.  This is consistent with the findings of \citet{rebull07} for the
Perseus 24\micron\ sources
and  by \citet{har07} for the Serpens IRAC sources.
Although there are undoubtedly some variable sources in these clouds, the observing
techniques of the c2d program were not designed to enable reliable detection of
modestly variable objects.

\subsection{Source Counts}\label{counts}

\begin{figure}
\epsscale{1.0}
\plotone{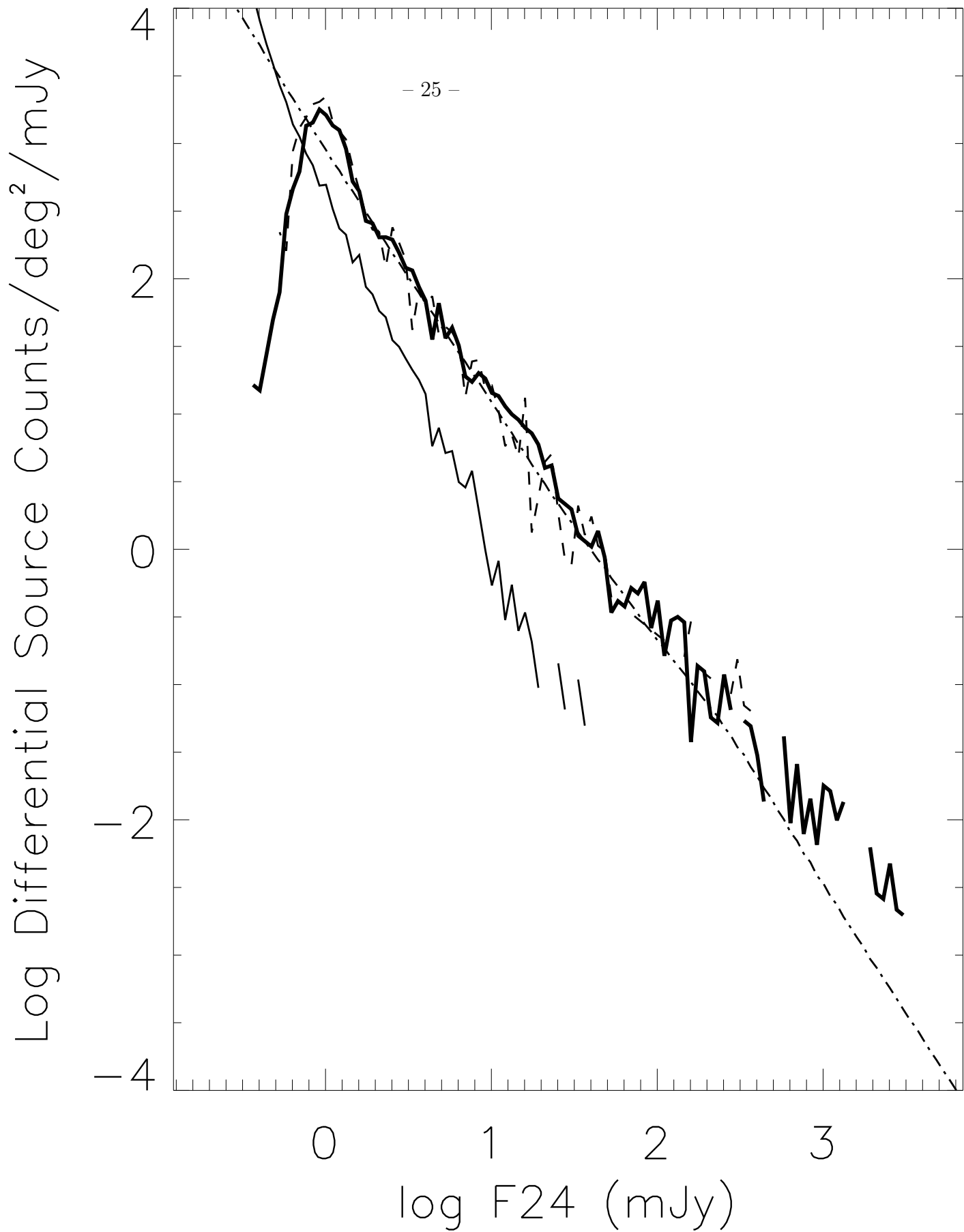}
\caption{24\micron\ source counts in the Serpens MIPS field (dark line), and off-cloud
region (dashed line). SWIRE galaxy
counts (thin line) fall below the Serpens data at our flux limit of 1 mJy.
The predicted source counts
from the Wainscoat model at 25 $\mu$m \citep{wains92} are shown by the dot-dash line.
\label{diffcnt}}
\end{figure}

Because the Serpens star-forming cloud is so close to the Galactic plane, {\it b} $\sim$ 5 degrees,
the vast majority of the sources detected at the shorter wavelengths are background
stars in the Galaxy.  At tfainter flux levels, background extragalactic objects constitute
a significant population.
In order to estimate the background Galactic star numbers we have used the \citet{wains92}
model provided by J. Carpenter (private communication).   Figure \ref{diffcnt} shows
the predicted star counts from the model together with the observed counts at 24\micron\ for
both the Serpens Cloud and the off-cloud region.  Also shown in the figure are the source counts
from the c2d-processed SWIRE ELAIS N1 field which are largely extragalactic for fluxes below
a few mJy.  This figure shows that contamination by Galactic stars at the brighter fluxes and by
extragalactic sources at the faint end is a significant problem for identifying Serpens Cloud
members.  To address this problem we discuss our use of several color and flux criteria in
the following section.  It is also apparent that there is an excess of bright ($F >$ 300 mJy)
sources relative to the expected background counts.  This excess is, in fact, real and represents
the bright end of the YSO candidate population discussed in the following section.

\subsection{Color-Magnitude Diagrams}\label{color}

\begin{figure}
\epsscale{1.0}
\plotone{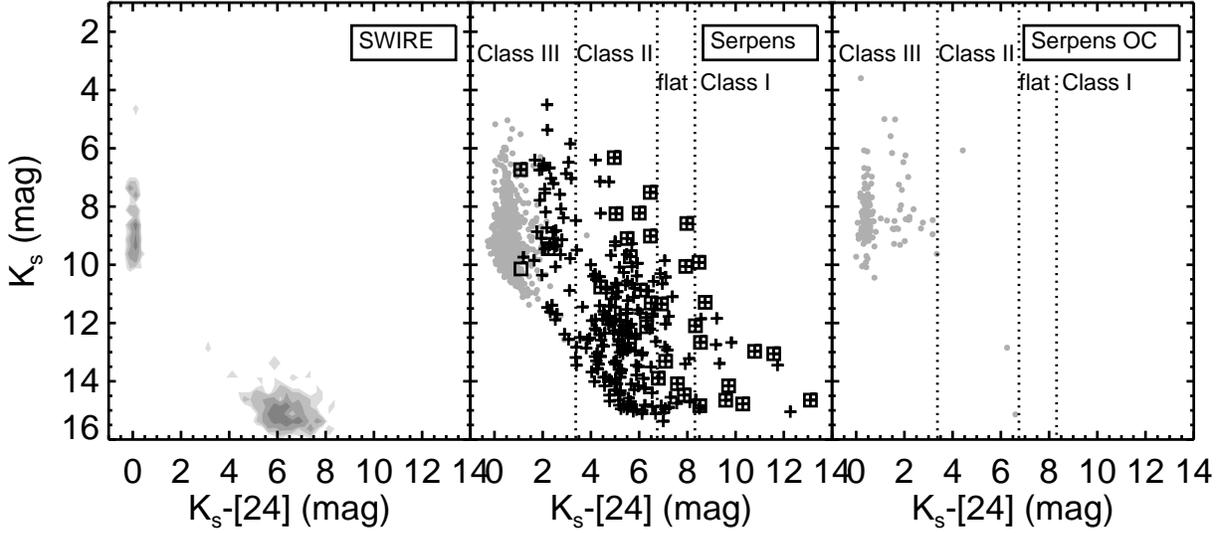}
\caption{
Color-magnitude diagram for $K_s$ vs.\ $K_s-[24]$ for objects in SWIRE (left) and Serpens
(center) and off-cloud region (right).  The SWIRE counts are shown as a surface density
with darker implying higher density.  Objects in SWIRE are expected to be mostly
galaxies (objects with $K_s\gtrsim$14) or stellar photospheres
(objects with $K_s-[24]\lesssim$1).  For the Serpens and off-cloud plots, 
filled gray circles are objects with SEDs resembling
photospheres, and plus signs are the remaining objects.  An
additional box around a point denotes that it was also detected
at 70\micron.  Objects that are candidate young objects have
colors unlike those objects found in SWIRE, e.g.,
$K_s\lesssim$14 and $K_s-[24]\gtrsim$1. Dashed lines denote the
divisions between Class I, flat, Class II, and Class III
objects; to omit foreground and background stars, we have
further imposed a $K_s-[24]>$2 requirement on our Class III
objects (see text).
\label{fig_k24a}}
\end{figure}

The c2d team has discussed in a number of studies how the use of color-magnitude  and
color-color diagrams can separate likely young cloud members with infrared excesses from
reddened stars and many background extragalactic sources \citep{young05,har06,rebull07,har07}.
Since nearly half of the area covered by our MIPS 24\micron\ observations was not
observed with IRAC \citep{har06}, we utilize the color and magnitude criteria developed
by \citet{young05} and refined by \citet{rebull07} and \citet{chapman07} to isolate a candidate YSO population
without requiring the existence of IRAC data.
The most populated diagram is naturally the color-magnitude diagram of K$_s$ versus
K$_s$ - [24] because of the much larger number of 24\micron\ sources than 70\micron\ ones.
Figure \ref{fig_k24a} shows the distribution of sources in this diagram for the 1453  
sources with S/N above 5 at 24\micron\ and with 2MASS K$_s$ matches within 4".
This distribution is very similar to that seen in other well-populated c2d clouds such
as Perseus \citep{rebull07}.
A comparison of the SWIRE results, the Serpens off-cloud results, and the Serpens Cloud
data shows: 1) objects in our ``star'' class fall in a relatively narrow band with
blue K$_s$-[24] colors (K$_s$-[24] $< 1$) as would be expected, and 2)  the part of the diagram toward redder colors
is populated by a number of sources in Serpens that are not seen in either the off-cloud
region or in the SWIRE data set, except at K magnitudes fainter than K$_s \sim$ 14.
This allows us to assign a high probability that sources in the region K$_s <$ 14 and
K$_s$ -[24] $>$ 2 are Serpens Cloud YSO candidates with excess emission at 24\micron\ probably
due to circumstellar dust.  Note that the off-cloud area does have a population of moderately
reddened objects (K$_s$-[24] $< 2$), well-fit as stellar photospheres that are not seen in the SWIRE sample, simply
because even the off-cloud area has more reddening than the high Galactic latitude ELAIS N1 region.
In order to categorize our YSO candidates crudely in terms of evolutionary state, we
have drawn lines in Figure \ref{fig_k24a} indicating where objects would fall based
on the YSO source classification criteria of \citet{greene94} using the K$_s$-[24] color to
measure the spectral slope.
Table \ref{yso-counts} lists the number of candidates and the number in each of the four classes.
Although AGB stars with substantial mass loss also exhibit
mid-infrared excesses, \citet{har06} have argued that the number expected in this
area is less than or of order a half dozen (four of which have already been confirmed spectroscopically
as AGB interlopers by Merin et al. (in prep.).  The positions of and photometry for the YSO
candidates that are not in the area covered by IRAC are given by \citet{har07} along with
those in the IRAC area.

\citet{har07} discuss the comparison between YSO's selected by the criteria used here (K$_s$ and
24\micron\ data only) and the more restrictive criteria possible with the combination of IRAC
data.  They basically find that we actually may have missed 8 or 9 YSO's in the area not covered by
IRAC and included a very few, 3 or 4, that may be background extragalactic sources.  But the overall
conclusion is that there is a good correspondence between the YSO candidates found using only
MIPS and 2MASS versus those selected with a more complete data set.  It is also clear that the
area mapped by both IRAC and MIPS, 0.85 deg$^2$ contains a much higher density of YSO's, 235 or
276 deg$^{-2}$ than does the area only covered by MIPS/24\micron\ with 51 YSO's or 54 deg$^{-2}$.
Even if we exclude the area of the two high density clusters, the area covered by the combined
IRAC/MIPS observations has a YSO density a factor of 4 higher than the area not included in
the IRAC observations.

\begin{figure}
\epsscale{1.0}
\plotone{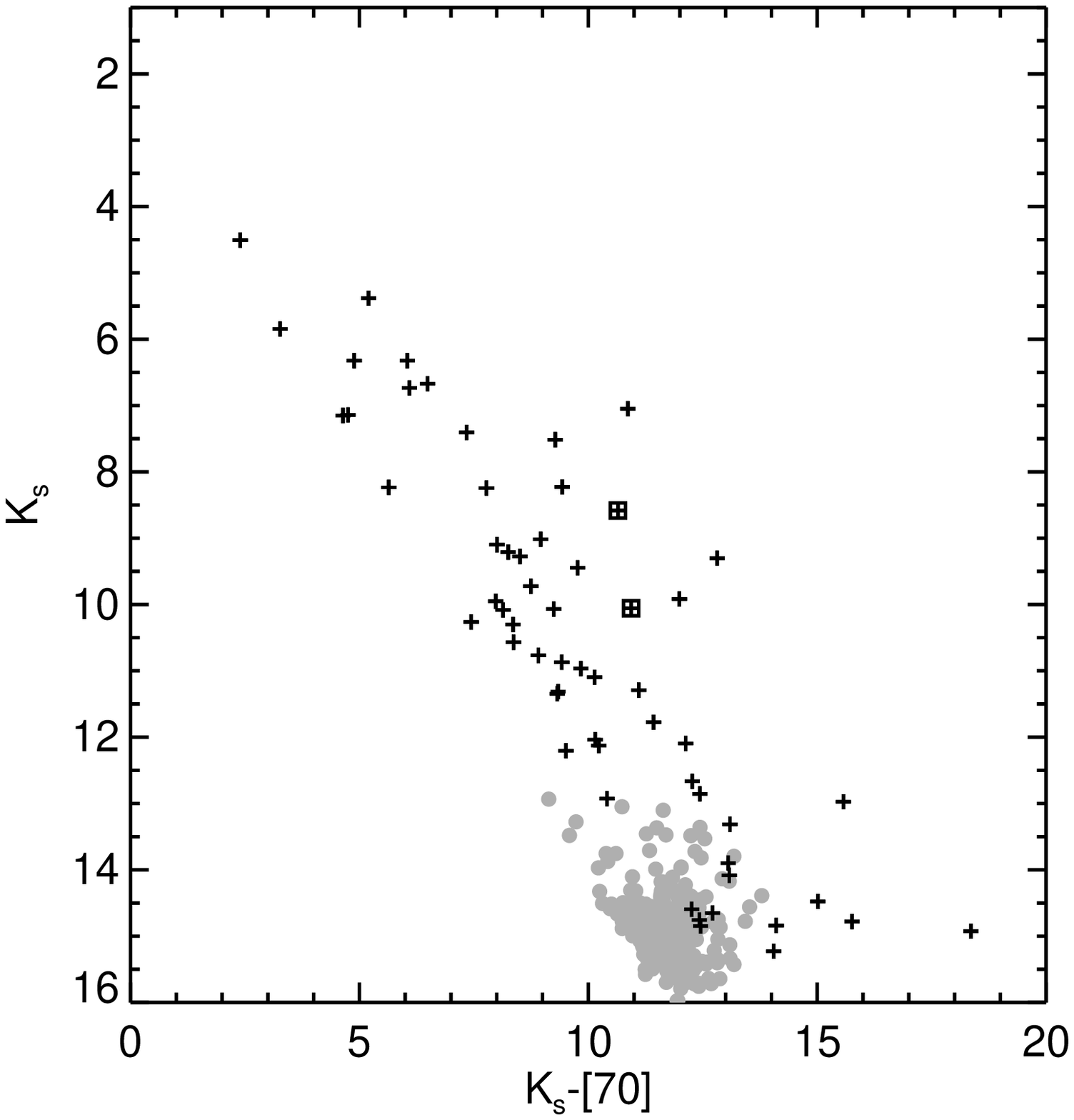}
\caption{Color-magnitude diagram of $K_s$ vs. $K_s - [70]$ for Serpens (crosses) with
data from the full SWIRE survey (grey dots) included for comparison.
\label{fig_cmK70}}
\end{figure}

We have also plotted our photometry in two other color/magnitude spaces for comparison with
other c2d clouds.  Figure \ref{fig_cmK70} shows the distribution of sources in K$_s$ vs. K$_s$-[70]
space.   As observed by \citet{rebull07} in Perseus, there are a large number of likely
cloud members at much brighter K$_s$ magnitudes than seen for SWIRE extragalactic
objects.  In addition, there is a small population of faint (in K$_s$) objects that are
redder than any of the SWIRE objects in both Serpens and Perseus.  The four objects
redder than K$_s$-[70] = 15 are all likely to be slightly less extreme versions of the
sources discussed in the next section.  Two of these are located in cluster A, but
tend to be around the outside of the tight cluster of very red objects.  The other two are in a small grouping
associated with 
the second of the four 160\micron\ point sources
listed in Table \ref{tbl_160}.  Since all of these objects were also observed in our program with
IRAC, they are also listed in the appropriate tables of \citet{har07}, and all are considered
high probability YSO's.

\begin{figure}
\epsscale{1.0}
\plotone{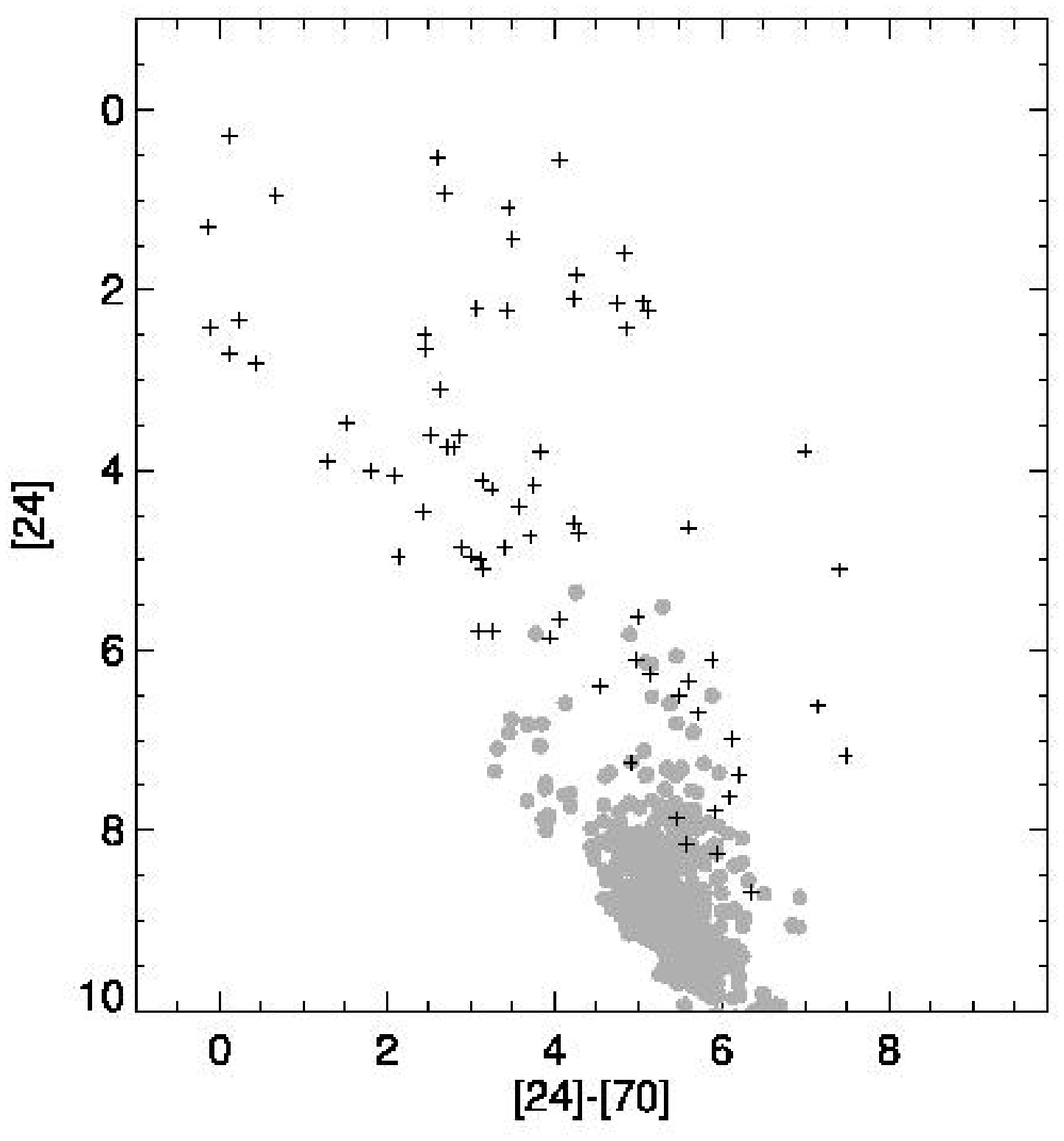}
\caption{Color-magnitude diagram of $[24]$ vs. $[24] - [70]$ for Serpens (crosses) with
data from the full SWIRE survey (grey dots) included for comparison.
\label{fig_cm2470}}
\end{figure}

The final color-magnitude diagram, [24] vs [24]-[70] is shown in Figure \ref{fig_cm2470}.
Again this distribution is qualitatively similar to that in Perseus, although we find many fewer
sources in the area overlapping the red edge of the extragalactic distribution than did 
\citet{rebull07} for their ``rest of the cloud''.  The Serpens distribution is qualitatively more
similar to that for the NGC1333 portion of Perseus.  Since many of the sources represented in
this diagram for Serpens are located in one of the two principal clusters, A and B, in Serpens,
it is perhaps not surprising that they would mimic some of the properties of similar young clusters
like NGC 1333.

\subsection{The Most Embedded Objects}\label{embed}

\begin{figure}
\plotfiddle{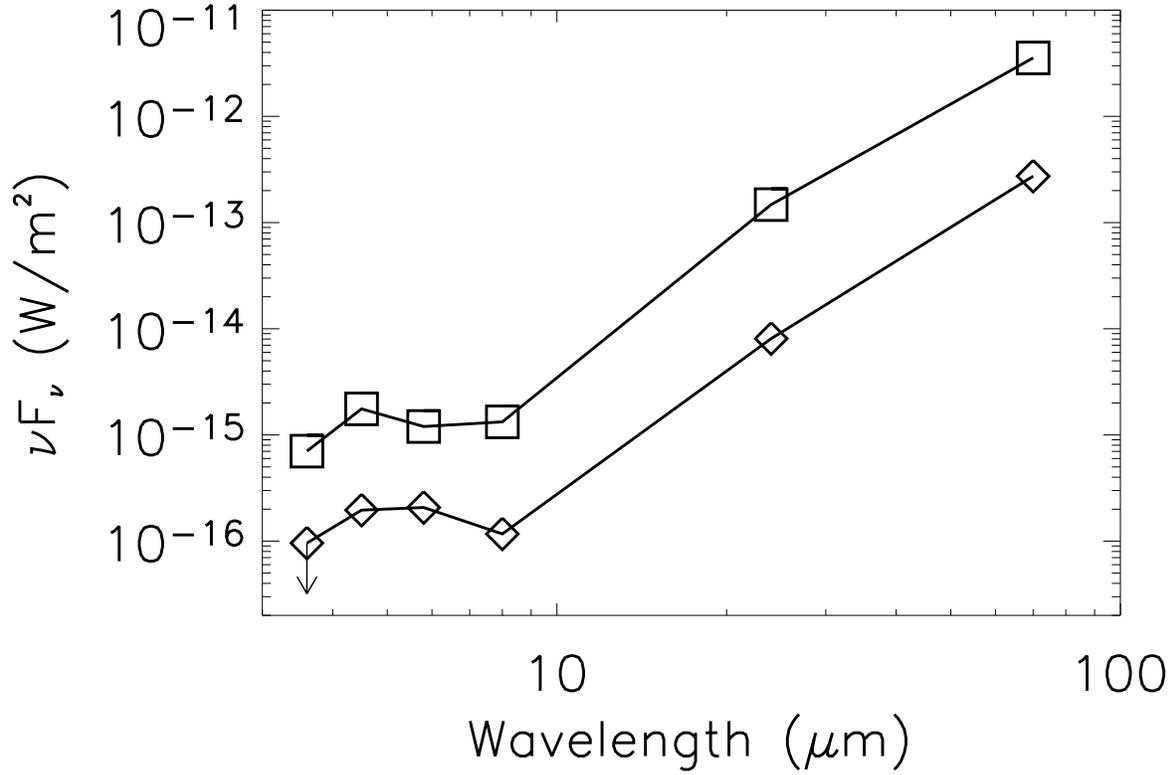}{5.0in}{0}{95}{95}{-300}{-300}
\caption{Spectral energy distribution for the two most embedded sources in Table \ref{cold-table},
one in Cluster A (open squares, SSTc2dJ1829463+0115219) and one in Cluster B (open diamonds, SSTc2dJ18290906+0031323, source ``C'' in Table \ref{hh-table}), both of which
appear to be associated with outflows.
\label{sed-fig}}
\end{figure}

We have selected the coldest, most obscured sources from our sample by looking for
objects not detected in the 2MASS survey but detected with reasonable signal-to-noise
at both 24 and 70\micron.  There are 11 such objects in our surveyed area, and these are
listed in Table \ref{cold-table}.  Interestingly all 11 are located in the heart of either
Cluster A or B.  Additionally, as shown in Table \ref{cold-table} all were detected in
some or all IRAC bands.  Their energy distributions are all consistent with a designation
of Class I even though they are not included in Figure \ref{fig_k24a} since they were not
detected in the 2MASS survey.  In fact, several of these objects are strongly enough peaked
in the far-infrared that they have energy distributions consistent with some nominal
Class 0 sources despite the fact that all were detected with IRAC.  The class status of these
will be discussed further using mm data by Enoch et al. (2007, in prep.).
Figure \ref{sed-fig} shows the SED's for the two most embedded objects from Table \ref{cold-table}.
Each of these appears to be associated with an outflow in its respective cluster, and both
have very similar SED's that differ only in their absolute flux level by a factor of $\sim$ 10.

\begin{figure}
\plotfiddle{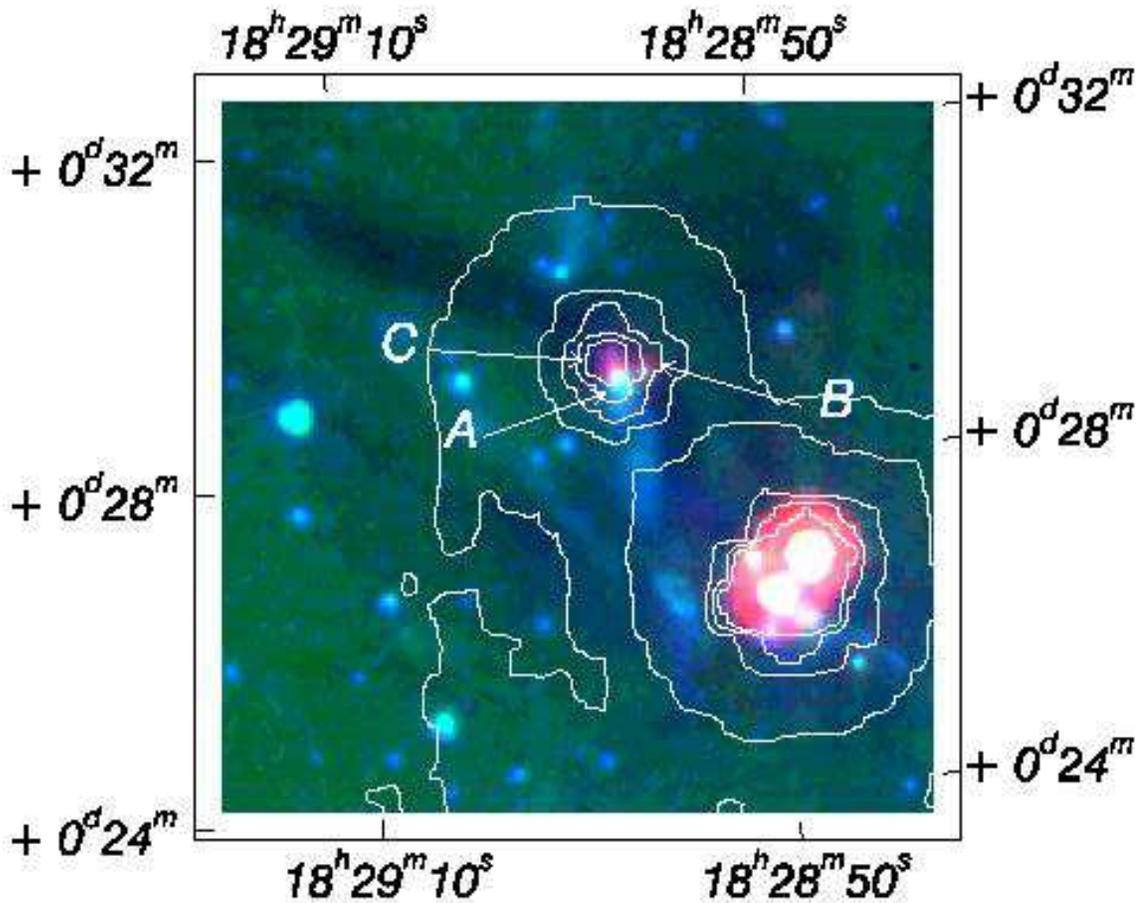}{7.0in}{-90}{95}{95}{-350}{400}
\caption{Three color image of the eastern end of Cluster B where the most embedded source, C, is
located.  This is the likely exciting source for an HH-like outflow visible in the IRAC data.
The color scheme is: blue/4.5\micron, green/8.0\micron, and red/24\micron.  The contours of
70\micron\ emission are also superimposed with levels at 40, 80, 160, 240, and 320 MJy/sr.
Also shown are the positions of two other compact sources extracted from the images in this
region.  The letters correspond to positions/fluxes in Table \ref{hh-table}.
\label{hh-fig}}
\end{figure}

Table \ref{cold-table} shows also that the most embedded object in Cluster B (whose SED is
shown in Figure \ref{sed-fig}) was not selected
as a YSO by \citet{har07}.  The reason is that the flux at 3.6\micron\ was too faint to meet
the selection criteria of that study.  The area within 15" of that source contains two other
extracted compact sources in the c2d data set.  The positions and photometry for all three
are shown in Table \ref{hh-table} and an image of the area is shown in Figure \ref{hh-fig}.
Although the source density is quite high, the 70\micron\ contours shown in the figure are
clearly centered on the northernmost source, ``C''.  Source ``B'' is a slightly extended source
that may represent a separate exciting object or may just be the location of the most visible
jet emission that has been discussed briefly by \citet{har07} in this region.  Source ``A'' is
a faint, but very red object about 6" to the west of source ``C'' and appears to be a
point-like object in the images.

Figure \ref{fig_avimage} shows clearly that Cluster A and B are located in the highest extinction parts of
the cloud.  Therefore the lack of detection of the objects in Table \ref{cold-table} at
1 -- 2.3\micron\ may be due at least partly to the extinction of the cloud material in
which they are embedded in addition to individual circumstellar material.  Although the
nominal extinction values in these areas range up to A$_v \sim$ 35 -- 40, the fact that
these values result from smoothing over 90 arcseconds of the stellar distribution means that
they probably underestimate the extinction in the most extreme regions.
This association of the coldest objects with the highest extinction regions is similar to
the correlation seen by \citet{enoch07} between extinction and location of dense mm cores.

\section{Summary}

We have described the basic observational characteristics of the c2d MIPS observations
of the Serpens Cloud.   In a 1.5 deg$^2$ area we have found 250 YSO candidates on the basis
of the K$_s$-[24] color.  An additional 11 objects can be identified on the basis of their
24 and 70\micron\ fluxes and lack of detection by 2MASS.  All of these YSO candidates will be
discussed in more detail in a companion paper \citep{har07}.  All the most embedded objects
are found in the central area of the two main clusters of YSO's previously identified in
Serpens.  The images and source catalogs derived from these data are all available on the
SSC website, http://ssc.spitzer.caltech.edu/legacy/all.html.

\acknowledgments
Support for this work, part of the Spitzer Legacy Science
Program, was provided by NASA through contracts 1224608, 1230782, and
1230779 issued by the Jet Propulsion Laboratory, California Institute
of Technology, under NASA contract 1407. 
Astrochemistry in Leiden is supported by a NWO Spinoza grant
and a NOVA grant. JKJ was supported by NASA Origins grant NAG5-13050.
This publication makes use of data products from the
Two Micron All Sky Survey, which is a joint project of the University
of Massachusetts and the Infrared Processing and Analysis
Center/California Institute of Technology, funded by NASA and the
National Science Foundation.  We also acknowledge extensive use of the SIMBAD
data base.

\clearpage

\begin{deluxetable}{lccccccccc}
\tablecolumns{10}
\tablecaption{Summary of Observations\label{table_summary}}
\tablewidth{0pt}
\tablehead{
\colhead{Region}      &   \colhead{AOR}   & \colhead{Time-Date}        & \colhead{{\it l}}\tablenotemark{a}&\colhead{{\it b}}\tablenotemark{a}\\
            &         &   \colhead{(UT)}           & \colhead{(deg)} & \colhead{(deg)} \\
}
\startdata
Serpens & 5713408 & 2004-04-05 23:40 & 31.5 & 5.4 \cr
            & 5713920 & 2004-04-06 04:05 & 31.5 & 5.3 \cr
            & 5713664 & 2004-04-06 00:22 & 31.6 & 5.2 \cr
            & 5714176 & 2004-04-06 04:48 & 30.6 & 5.1 \cr
Off Cloud & 5716736 & 2004-04-06 01:26 & 35.2 & 4.4 \cr
            & 5716992 & 2004-04-06 05:52 & 35.2 & 4.3 \cr
\enddata
\tablenotetext{a}{\ {\it l} and {\it b} are listed for the center of the 24 $\mu$m AOR.}
\end{deluxetable}

\clearpage

\begin{deluxetable}{lccc}
\tablecolumns{4}
\tablecaption{Serpens Cloud Sky Coverage\label{table_coverage}}
\tablewidth{0pt}
\tablehead{
\colhead{Region}      & \colhead{24 $\mu$m} &  \colhead{70 $\mu$m} & \colhead{160 $\mu$m} \\
            & \colhead{(deg$^2$)} &  \colhead{(deg$^2$)} & \colhead{(deg$^2$)}\\ 
}
\startdata
Serpens          & 1.81 & 1.57 & 1.49 \cr
Off-Cloud            & 0.47 & 0.36 & 0.41 \cr
\enddata
\end{deluxetable}
\clearpage

\begin{deluxetable}{ccccc}
\tablecolumns{5}
\tablecaption{160\micron\ Point Sources\label{tbl_160}}
\tablewidth{0pt}
\tablehead{
\colhead{RA (J200)}   & \colhead{Dec (J200)} & \colhead{Flux (mJy)} & \colhead{Comment} & \colhead{YSO\#}\tablenotemark{a} \\
}
\startdata
18 29 32.3  & $+$01 18 56  & 24000 & Single 24/70\micron\ Source & 104\cr
18 29 52.9  & $+$00 36 09  & 18200 & Cluster of four 24\micron\ Sources\cr
18 29 16.7  & $+$00 18 20  & 10000 & Single 24/70\micron\ Source & 88\cr
18 28 15.7  & $-$00 03 11  & 6070 & Cluster of four 24\micron\ Sources\cr
\enddata
\tablenotetext{a}{YSO number from \citet{har07}.}
\end{deluxetable}
\clearpage

\begin{deluxetable}{lc}
\tablecolumns{2}
\tablecaption{Serpens Cloud Detection Statistics\label{src-counts}}
\tablewidth{0pt}
\tablehead{
\colhead{Wavelength(s)}      & \colhead{Source Number} \\
}
\startdata
24\micron\ $>3\sigma$   & 2635 \cr
24\micron\ $>5\sigma$   & 1494 \cr
70\micron\ $>3\sigma$   & 97 \cr
70\micron\ $>5\sigma$   & 88 \cr
24 \& 70\micron\ $>5\sigma$   & 75 \cr
24\micron\ \& 2MASS K$_s$ $>5\sigma$   & 1085 \cr
24\micron\ \& any IRAC   & 1040\tablenotemark{a} \cr
70\micron\ \& any IRAC   & 77 \cr
\enddata
\tablenotetext{a}{The greater number of matches between 24\micron\ and K$_s$ versus IRAC is due to the smaller area coverage of the IRAC data.}
\end{deluxetable}
\clearpage

\begin{deluxetable}{lc}
\tablecolumns{2}
\tablecaption{Classification based on \ks$-$[24] \label{yso-counts}}
\tablewidth{0pt}
\tablehead{
\colhead{Classification} &  \colhead{Serpens Source Count\tablenotemark{a}} }
\startdata
number with \ks$-$[24]$>$2, \ks$<$14 & 250 \\
number with \ks$-$[24]$>$2, \ks$<$14, and Class I \ks$-$[24] color
& 15 (6\%) \\
number with \ks$-$[24]$>$2, \ks$<$14, and ``flat'' \ks$-$[24]
color & 21 (8\%)  \\
number with \ks$-$[24]$>$2, \ks$<$14, and Class II \ks$-$[24]
color & 158 (63\%) \\
number with \ks$-$[24]$>$2, \ks$<$14, and Class III \ks$-$[24]
color & 56 (22\%)\\
\enddata
\tablenotetext{a}{Since a 2MASS detection is required to be included in these statistics, very cold or deeply embedded sources
are not present in these counts, e.g. those sources in Table \ref{cold-table}.}
\end{deluxetable}

\begin{deluxetable}{lcccccccc}
\tabletypesize{\small}
\rotate
\tablecolumns{9}
\tablecaption{The Most Embedded Objects \label{cold-table}}
\tablewidth{0pc}
\tablehead{
\colhead{Name/Position} &
\colhead{YSO \#}\tablenotemark{a} &
\colhead{3.6 \micron\ }  &
\colhead{4.5 \micron\ }  &
\colhead{5.8 \micron\ }  &
\colhead{8.0 \micron\ }  &
\colhead{24.0 \micron\ }  &
\colhead{70.0 \micron\ } &
\colhead{Associated Source}\tablenotemark{b}        \\
\colhead{SSTc2dJ...} &
\colhead{}      &
\colhead{(mJy)}  &
\colhead{(mJy)}  &
\colhead{(mJy)}  &
\colhead{(mJy)}  &
\colhead{(mJy)}  &
\colhead{(mJy)}  &
}
\startdata
18285404$+$0029299 & 40  & 5.81$\pm$0.50 & 27.6$\pm$ 2.3 & 44.8$\pm$ 2.6 & 56.4$\pm$ 3.2 &  918$\pm$  85 & 11100$\pm$  1040 & D62/66 \\
18285486$+$0029525 & 42  & 1.94$\pm$0.12 & 10.6$\pm$ 0.6 & 20.4$\pm$ 1.1 & 30.2$\pm$ 1.6 &  765$\pm$  70 & 7250$\pm$  675 & D65\\
18290619$+$0030432 & 67  & 8.05$\pm$0.41 & 45.0$\pm$ 2.8 & 93.9$\pm$ 4.8 &  129$\pm$   7 & 1320$\pm$  139 & 7240$\pm$  713 & D90\\
18290675$+$0030343 & 68  & 3.27$\pm$0.21 & 11.7$\pm$ 0.7 & 14.9$\pm$ 0.8 & 20.7$\pm$ 1.2 & 1000$\pm$  105 & 11400$\pm$  1180& D94\\
18290906$+$0031323 &  & $<$  0.12 & 0.29$\pm$0.03 & 0.40$\pm$0.09 & 0.31$\pm$0.08 & 64.6$\pm$ 6.0 & 6380$\pm$  611 & D101\\
18294810$+$0116449 & 135  & 1.96$\pm$0.10 & 6.98$\pm$0.42 & 12.1$\pm$ 0.6 & 16.7$\pm$ 0.8 &  219$\pm$  21 & 14900$\pm$  1420 & K241, SMM9\\
18294963$+$0115219 & 141  & 0.85$\pm$0.08 & 2.64$\pm$0.27 & 2.32$\pm$0.28 & 3.54$\pm$0.31 & 1180$\pm$  117 & 82800$\pm$  7810 & K258a, SMM1\\
18295219$+$0115478 & 150  & 7.38$\pm$0.41 & 33.0$\pm$ 2.1 & 41.3$\pm$ 2.2 & 40.0$\pm$ 2.6 & 1640$\pm$  154 & 15200$\pm$  1420 & K270, SMM10\\
18295285$+$0114560 & 155  & 8.65$\pm$0.44 & 34.6$\pm$ 1.8 & 72.0$\pm$ 3.4 &  110$\pm$   5 & 1040$\pm$   96 & 5570$\pm$  523 & K276\\
18295927$+$0114016 & 195  & 2.72$\pm$0.28 & 5.76$\pm$0.44 & 7.78$\pm$1.16 & 36.0$\pm$ 5.4 &  109$\pm$  19 & 12200$\pm$  1160 & SMM3\\
18295992$+$0113116 & 198  & 2.77$\pm$0.16 & 29.5$\pm$ 1.5 &  103$\pm$   4 &  199$\pm$  10 & 2620$\pm$  249 & 6830$\pm$  675 & K331\\

\enddata
\tablenotetext{a}{Identifying number from YSO table in \citet{har07}.}
\tablenotetext{b}{References are: D: \citep{djup06}, K: \citep{kaas04}, SMM: \citet{dav99}.}
\end{deluxetable}

\begin{deluxetable}{llccccccc}
\tabletypesize{\small}
\rotate
\tablecolumns{9}
\tablecaption{Sources Marked In Figure \ref{hh-fig} \label{hh-table}}
\tablewidth{0pc}
\tablehead{
\colhead{Marker} &
\colhead{Name/Position} &
\colhead{3.6 \micron\ }  &
\colhead{4.5 \micron\ }  &
\colhead{5.8 \micron\ }  &
\colhead{8.0 \micron\ }  &
\colhead{24.0 \micron\ }  &
\colhead{70.0 \micron\ } \\
\colhead{} \\
\colhead{SSTc2dJ...} &
\colhead{(mJy)}  &
\colhead{(mJy)}  &
\colhead{(mJy)}  &
\colhead{(mJy)}  &
\colhead{(mJy)}  &
\colhead{(mJy)}  &
}
\startdata
A\tablenotemark{a} & 18290904$+$0031280   &  0.95$\pm$0.11 & 2.78$\pm$0.23 & 2.92$\pm$0.24 & 5.03$\pm$0.40 & 14.0$\pm$ 1.9 & \nodata \\
B & 18290864$+$0031305   &  0.06$\pm$0.03 & 0.32$\pm$0.02 & 0.47$\pm$0.05 & 0.62$\pm$0.07 & 36.2$\pm$ 3.4 & \nodata \\
C & 18290906$+$0031323   & $<$  0.12 & 0.29$\pm$0.03 & 0.40$\pm$0.09 & 0.31$\pm$0.08 & 64.6$\pm$ 6.0 & 6380$\pm$  611 \\

\enddata
\tablenotetext{a}{This is YSO \# 75 in \citet{har07}.}
\end{deluxetable}

\clearpage

\clearpage

\clearpage
\begin{figure}
\plotfiddle{f3.e.ps}{5.0in}{90}{75}{75}{320}{0}
\figcaption{Completeness test at 24\micron.  The upper solid line shows the measured completeness
fraction for artificial sources inserted into the observed 24\micron\ mosaic image of Serpens
as a function of magnitude.  The slightly higher dash-dot line shows the completeness fraction
for sources inserted into an artificial image with no real sources but with a noise level
equal to that in the observed data.  The lower solid line (mostly equal to zero) shows the
fraction of ``unreliable'' sources, i.e. sources extracted which were not real. 
\label{complete}}
\end{figure}

\clearpage

\clearpage

\clearpage

\clearpage

\clearpage

\clearpage

\clearpage

\clearpage

\end{document}